\documentclass[10pt,twoside,twocolumn]{article}
\usepackage[utf8]{inputenc}
\usepackage[T1]{fontenc}
\usepackage{microtype}

\usepackage[super,sort&compress,comma]{natbib} 
\usepackage{times}
\usepackage{sectsty}
\usepackage{balance} 

\usepackage{graphicx} 
\usepackage{lastpage}
\usepackage[format=plain,justification=raggedright,singlelinecheck=false,font=small,labelfont=bf,labelsep=space]{caption} 
\usepackage{fancyhdr}
\usepackage{microtype}

\usepackage{color}
\usepackage[normalem]{ulem}
\definecolor{Red}{rgb}{0.9,0.0,0.1}
\definecolor{Lila}{rgb}{0.7,0.0,0.9}
\definecolor{Darkblue}{rgb}{0.22,0.33,0.64}
\definecolor{Darkgray}{rgb}{0.4,0.4,0.4}
\definecolor{Blue}{rgb}{0.1,0.0,0.9}

\usepackage{amsmath}
\usepackage{amssymb}
\usepackage{amsfonts}
\usepackage{units}
\usepackage{icomma}
\renewcommand {\vec} {\boldsymbol}                
\newcommand {\diff} {\mathrm{d}}                 
\newcommand {\EHo} {E_\text{2D}}                 
\newcommand {\FvK} {\gamma_\text{FvK}}
\newcommand {\order}[1] {{\mathcal O}\!\left( #1 \right)}
\newcommand {\qq}[1] {\quad \text{#1} \quad }   
\newcommand {\zweiervec}[2] {\begin{pmatrix}#1\\#2\end{pmatrix}}

\let\phi\varphi                     

\let\epsilon\varepsilon

\let\Psi\varPsi

\begin{document}

\thispagestyle{plain}
\fancypagestyle{plain}{
\renewcommand{\headrulewidth}{1pt}}
\renewcommand{\thefootnote}{\fnsymbol{footnote}}
\renewcommand\footnoterule{\vspace*{1pt}%
\hrule width 3.4in height 0.4pt \vspace*{5pt}} 
\setcounter{secnumdepth}{5}

\makeatletter 
\def\subsubsection{\@startsection{subsubsection}{3}{10pt}{-1.25ex plus -1ex minus -.1ex}{0ex plus 0ex}{\normalsize\bf}} 
\def\paragraph{\@startsection{paragraph}{4}{10pt}{-1.25ex plus -1ex minus -.1ex}{0ex plus 0ex}{\normalsize\textit}} 
\renewcommand\@biblabel[1]{#1}            
\renewcommand\@makefntext[1]%
{\noindent\makebox[0pt][r]{\@thefnmark\,}#1}
\makeatother 
\renewcommand{\figurename}{\small{Fig.}~}
\sectionfont{\large}
\subsectionfont{\normalsize} 

\fancyhead{}
\renewcommand{\headrulewidth}{1pt} 
\renewcommand{\footrulewidth}{1pt}
\setlength{\arrayrulewidth}{1pt}
\setlength{\columnsep}{6.5mm}
\setlength\bibsep{1pt}

\twocolumn[
  \begin{@twocolumnfalse}
\noindent\LARGE{\textbf{Osmotic buckling of spherical capsules}}
\vspace{0.6cm}

\noindent\large{\textbf{Sebastian Knoche, and
Jan Kierfeld$^{\ast}$}}\vspace{0.5cm}

\vspace{0.6cm}

\noindent \normalsize{
We study the buckling of elastic spherical shells under osmotic pressure with the osmolyte concentration of the exterior solution as control parameter. We compare our results for the bifurcation behavior with results for buckling under mechanical pressure control, that is, with an empty capsule interior. We find striking differences for the buckling states between osmotic and mechanical buckling. Mechanical pressure control always leads to fully collapsed states with opposite sides in contact, whereas uncollapsed states with a single finite dimple are generic for osmotic pressure control. For sufficiently large interior osmolyte concentrations, osmotic pressure control is qualititatively similar to buckling under volume control with the volume prescribed by the osmolyte concentrations inside and outside the shell. We present a quantitative theory which also captures the influence of shell elasticity on the relation between osmotic pressure and volume. These findings are relevant for the control of buckled shapes in applications. We show how the osmolyte concentration can be used to control the volume of buckled shells. An accurate analytic formula is derived for the relation between the osmotic pressure, the elastic moduli and the volume of buckled capsules. This also allows to use elastic capsules as osmotic pressure sensors or to deduce elastic properties and the internal osmolyte concentration from shape changes in response to osmotic pressure changes. We apply our findings to published experimental data on polyelectrolyte capsules. } \vspace{0.5cm}
 \end{@twocolumnfalse}
  ]

\section{Introduction}
\footnotetext{\textit{TU Dortmund University, 
    Department of Physics, 44221 Dortmund. E-mail:
    Jan.Kierfeld@tu-dortmund.de}}

Elastic capsules consist of an elastic spherical shell enclosing a fluid
phase. They are commonly met in nature, prominent examples exhibiting elastic
properties similar to elastic shells are red blood cells\cite{Discher1998},
virus capsules\cite{Michel2006}, or pollen
grains\cite{Katifori2010}. Artificial capsules can be fabricated by various
methods,\cite{Meier2000,Fery2004,Neubauer2013} for example by interfacial
polymerization at liquid droplets\cite{Rehage2002} or by multilayer deposition
of polyelectrolytes\cite{Donath1998}, and have numerous applications as
delivery systems. Capsules are easily deformed by mechanical forces and their
deformation behavior exhibits buckling instabilities upon decreasing the
interior pressure or the enclosed volume.\cite{Gao2001, Okubo2001, Okubo2003,
  Fery2004, Zoldesi2008, Quilliet2008, Sacanna2011, Datta2012} These
deformation modes can potentially be used to infer material properties of the
enclosing shell material\cite{Knoche2011,Jose2014,Knoche2013,Neubauer2013} or
to control the shapes of capsules for applications\cite{Sacanna2011}.

Theoretically, the buckling instability of a spherical shell can be described
within classical shell theory,\cite{Ventsel2001, Landau1970, Hutchinson1967,
  Koiter1969, Timoshenko1959, Timoshenko1961} which identifies a critical
pressure where the spherical shape becomes unstable with respect to decreasing
volume and developing a finite dimple. Beyond the critical mechanical pressure
buckled shapes with a small dimple remain unstable with respect to further
spontaneous growth of the dimple\cite{Koiter1969,Landau1970} until opposite
sides get into contact, and the shells snap-through into a fully collapsed
state.\cite{Knoche2011}

Despite this theoretical prediction of a spontaneous snap-through into
a collapsed state for buckling under mechanical pressure, buckled shapes with
a {\em finite} dimple, \emph{i.e.}, without contact of opposite capsule sides,
are usually observed in microcapsule experiments performed under osmotic
pressure control.\cite{Gao2001, Fery2004, Sacanna2011, Datta2012} 
Buckling by
 osmotic pressure is  intimately related to buckling by
controled volume reduction because an applied 
  external osmotic pressure defines an osmotically preferred volume. 
 The  capsule volume can also be considered as fixed
  when it is filled with an incompressible fluid that cannot leave the
  capsule, or leaves the capsule on a very slow time scale like in dissolving
  or drying mechanisms.\cite{Okubo2001,
    Okubo2003,Pauchard2004,Zoldesi2008,Quilliet2008, Datta2010} In such
volume controled experiments, buckled shapes with finite dimples 
are also stable configurations.
This raises
the questions to what extent buckling under osmotic pressure control with
the osmolyte concentration of the exterior solution as control parameter
differs from buckling under mechanical pressure control, where we assume an
``empty'' capsule interior, and to what extent it differs 
from buckling under volume control with the  volume
   prescribed by the condition of equal interior 
and exterior osmolyte concentrations.
 A precise theoretical answer to these questions is
highly relevant for the control and analysis of buckled shapes in
 applications. Eventually,
the shape of osmotically buckled capsules can also be used to sense the
osmotic pressure and to deduce elastic material parameters 
based on quantitative theoretical modeling.

\section{Model for axisymmetric shells}

We analyze axisymmetric shapes by the use of non-linear shell theory
\cite{Pozrikidis2003,Libai1998} from which we can derive axisymmetric shape
equations.\cite{Knoche2011} Solutions of these equations can represent stable,
metastable, or even unstable capsules shapes. Shape transitions or
bifurcations between different axisymmetric solution branches can be
investigated using general results from bifurcation theory.\cite{Maddocks1987}
If a spherical shell develops a dimple, we naively expect relevant shapes to
be axisymmetric. However, non-axisymmetric shapes are relevant both at the
onset of the buckling instability\cite{Hutchinson1967, Koiter1969} and for
heavily deflated thin shells that undergo a secondary buckling
transition.\cite{Knoche2014, Knoche2014long, Quilliet2008, Quilliet2012, Vliegenthart2011}
Here, we aim for a classification of the transition from the spherical to
the axisymmetric buckled shape under osmotic pressure, 
under mechanical pressure, and under volume control.  
Our analysis will reveal important differences 
in the resulting buckling pathway between these three types of control.

\subsection{The elastic energy functional}

We start with an elastic shell that is spherical in its relaxed state, with a radius $R_0$. This shape can be parametrized in polar cylindrical coordinates $(r_0, z_0)$ by
\begin{equation}
 \vec r_0 (s_0) = \zweiervec{r_0(s_0)}{z_0(s_0)} =
    \zweiervec{R_0 \sin(s_0/R_0)}{- R_0 \cos(s_0/R_0)}
\end{equation} 
with an arc-length coordinate $s_0$. The shell is deformed by a normal pressure difference $p\equiv p_{\rm in} - p_{\rm ex}$, which is spatially constant across the whole shell. The sign convention is such  that $p<0$ if the shell is being deflated.

Nonlinear shell theory can be used to calculate the parametrization $\vec
r(s_0)$ of the deformed shape from which the strains and stresses in the shell
can be deduced. Appropriate shape equations have been introduced in ref.\ \citenum{Knoche2011}, to which  the reader is referred for the full mathematical treatment.

For the stability discussion that will be presented in the next sections, the
essential feature of the shape equations is that they can be derived from an
energy functional by calculus of variations. The elastic energy that is stored
in the deformed shell depends on the meridional and circumferential stretches
$\lambda_s$ and $\lambda_\phi$ and the bending strains $K_s = \lambda_s
\kappa_s - 1/R_0$ and $K_\phi = \lambda_\phi \kappa_\phi - 1/R_0$ which
measure the change of curvature in meridional and circumferential direction,
with $\kappa_s$ and $\kappa_\phi$ being the principal curvatures of the
deformed midsurface.\cite{Knoche2011,Libai1998} They can be calculated from
the parametrizations $\vec r_0(s_0)$ and $\vec r(s_0)$ of the reference shape
and deformed shape, respectively.

The surface energy density $w_S$ measures the elastic energy per undeformed area,\cite{Knoche2011,Libai1998}
\begin{multline}
 w_S = \frac{1}{2} \frac{\EHo}{1-\nu^2} 
 \big( [\lambda_s-1]^2 + 2 \nu [\lambda_s - 1][\lambda_\phi - 1] \\ 
   + [\lambda_\phi-1]^2 \big)
 + \frac{1}{2} E_B \left( K_s^2 + 2 \nu K_s K_\phi + K_\phi^2 \right)
\end{multline} 
with the two-dimensional Young modulus $\EHo$, the two-dimensional Poisson ratio $\nu$ and the bending stiffness $E_B$. For a  shell consisting of a thin sheet of isotropic material, these material constants are related to the bulk moduli  by $\EHo = E H_0$, $\nu = \nu_\text{3D}$ and $E_B = {E H_0^3}/({12(1-\nu^2)})$, where $H_0$ is the shell thickness, $E$ is the (three-dimensional) Young modulus and $\nu_\text{3D}$ is the (three-dimensional) Poisson ratio.

The elastic energy functional can now be written as the integral of the energy density over the undeformed shape with surface element $\diff A_0 = 2 \pi r_0 \diff s_0$,
\begin{equation}
 U[\vec r] = \int 2 \pi r_0 w_S \, \diff s_0.
\end{equation}

\subsection{Mechanical  pressure control and volume control}

In order to describe the deflation of the shell, additional terms must be
incorporated in the energy functional that account for the external
loads. When there is a prescribed mechanical pressure difference $p$ between
the inside and outside, the appropriate \emph{load potential} is 
$- p V[\vec r]$ 
where the volume $V = \int \pi r^2(s_0) z'(s_0) \, \diff s_0$
is a functional of the shape.\cite{Knoche2011,Libai1998} Then the shape
equations follow from minimizing the enthalpy functional $H[\vec r] = U[\vec
r] - p V[\vec r]$ for given $p$, which means that the first variation must
vanish,
\begin{equation} \label{eq:delta_H}
 \delta H = \delta U - p \delta V = 0,
\end{equation} 
see ref.\ \citenum{Knoche2011} for the resulting Euler-Lagrange equations.

This minimization can be interpreted in two ways: either as an unconstrained
minimization of the enthalpy functional $H$ or, alternatively, as a
minimization of the functional $U$ under the constraint that the functional
$V[\vec r]$ equals some given volume.  The pressure $p$ is then merely a
Lagrange multiplier to control the shell volume.
These two cases, termed \emph{(mechanical) pressure control} and \emph{volume
  control}, respectively, produce the same shapes as solutions of the shape
equations. However, the shapes show very different stability properties in the
two cases: While buckling under volume control will start with relatively
small (but finite) dimples and the size of the dimple is precisely controled
by the prescribed volume, buckling under mechanical pressure control will lead
to a complete collapse of the shell, so that opposite sides are in contact
with each other.\cite{Knoche2011} A detailed discussion follows below. Both
cases are idealized and hard to achieve in actual experiments: As long as
capsules are filled with some internal medium, there is a feedback between the
volume change and the internal pressure, so that the pressure difference $p$
is not fixed but varies with the capsule volume, which is in conflict with our
notion of pressure control. Also  in a typical volume control experiment, \emph{e.g.},
when an enclosed incompressible liquid evaporates, 
volume control is only an approximation whose quality depends on 
how large the  time scale for 
 evaporation is in comparison to the time scale for 
elastic shape relaxation.

\subsection{Osmotic pressure control}

Many deformation experiments with microcapsules 
are based on osmosis.\cite{Gao2001, Fery2004, Sacanna2011, Datta2012}
In osmotic buckling, solvent diffuses through the semi-permeable capsule
membrane because of an osmolyte concentration gradient between the inside and
outside. Osmosis tends to decrease the concentration gradient, and deflation
of the capsule stops when the concentrations in the inside and outside are
sufficiently matched. This is an important difference to mechanical pressure
control with an ``empty'' interior, where the deflation only stops when the
opposite sides of the capsule are in contact and the capsule volume is
virtually zero.

Ideal dilute solutions of osmotically active particles can be treated like
ideal gases.\cite{Callen1960} The appropriate energy functional that is to be
minimized in the case of osmosis must take into account the osmotic free
energy of the inner and outer solutions,\cite{Lipowsky2005}
\begin{equation}
  F_\text{os} = - k_B T N_\text{in} 
   \ln\left[ \frac{e}{\lambda_B^3} \frac{V}{N_\text{in}} \right] 
  - k_B T N_\text{ex} \ln\left[ \frac{e}{\lambda_B^3} 
   \frac{V_\text{ex}-V}{N_\text{ex}} \right].
\end{equation} 
In this expression, $k_B$ is the Boltzmann constant, $T$ the temperature of
the solutions, $\lambda_B = h/\sqrt{2 \pi m k_B T}$ the thermal de Broglie
wavelength with Planck constant $h$ and particle mass $m$.  $N_\text{in}$ and
$N_\text{ex}$ are the number of osmotically active particles inside and
outside the shell, respectively, $V$ the volume inside the shell and
$V_\text{ex} - V$ the outside volume. The osmotically active particles cannot
diffuse through the shell wall, such that $N_\text{in}$ is fixed during the
deflation; 
the experimental control parameter for osmotic pressure control 
is the number $N_\text{ex}$ of osmotically active molecules in the outside
solution via their concentration $N_\text{ex}/(V_\text{ex} - V)\approx
N_\text{ex}/V_\text{ex}$ (assuming $V \ll V_\text{ex}$).
 Furthermore, the temperature $T$ is considered to affect only the
ideal solutions; we do not incorporate thermal fluctuations in the elastic
shell, which is a good approximations unless shells are extremely 
thin.\cite{Paulose2012}

For $V \ll V_\text{ex}$, the second logarithm in $F_\text{os}$ can be
expanded 
and simplifies to
\begin{equation} \label{eq:F_os}
 F_\text{os} = - k_B T N_\text{in} \ln V 
    + \frac{k_B T N_\text{ex}}{V_\text{ex}} \, V + \text{const}.
\end{equation} 
Constant terms not depending on $V$ are fixed when we minimize the total
energy functional with respect to the shape of the shell, which only has an
influence on $V$ in eq.\ (\ref{eq:F_os}). The osmotic pressure difference can be
derived from this equation by
\begin{equation}\label{eq:p_os}
 p_\text{os} = - \partial{F_\text{os}}/\partial{V} = k_B T
(N_\text{in}/V - N_\text{ex}/V_\text{ex}) \equiv p_\text{in} - p_\text{ex}.
\end{equation}

The first term represents the internal osmotic pressure $p_\text{in}$, 
the second term 
  the external osmotic pressure $p_\text{ex}$, which also occurs in
eq.\ (\ref{eq:F_os}) as the prefactor of the term linear in $V$. 
The external pressure $p_\text{ex}$ is proportional to the 
external concentration of osmotically active particles, and thus it is
the experimentally controled pressure component.
The osmotic free
energy $F_\text{os}$ in eq.\ (\ref{eq:F_os}) is minimized by a volume $V =
N_\text{in} V_\text{ex}/N_\text{ex}$ indicating that the preferred state of
the system has equal concentrations of osmotically active particles inside and
outside the capsule.

The total energy functional accounts for the elastic energy of the deformed
shell and the free energies of the solutions, and reads
\begin{equation} \label{eq:G}
 G[\vec r] = U[\vec r] - k_B T N_\text{in} \ln V[\vec r] + p_\text{ex} V[\vec r].
\end{equation} 
In this functional, $U$ and $V$ depend on the shape of the shell, and its
variation is $\delta G = \delta U + (\partial{F_\text{os}}/\partial{V}) \delta V = \delta
U - p_\text{os} \delta V$. 
Thus, in comparison with mechanical pressure control 
as described by eq.\ (\ref{eq:delta_H}) and 
according to eq.\ (\ref{eq:p_os}), the
same shape equations are obtained with a  pressure
difference $p = p_\text{os} = p_\text{in} - p_\text{ex}$ exerted
on the shell.

Also for the experimental situation
 of a shell containing an ideal gas, the same energy functional (\ref{eq:G}) 
is obtained. The internal gas has a free energy $F_\text{gas} = -k_B T
N_\text{in} \ln V$, where $N_\text{in}$ is now the number of gas
atoms. According to the ideal gas equation $p V = N k_B T$, the prefactor can
also be written as $k_B T N_\text{in} = p_\text{in} V$ with an internal gas
pressure $p_\text{in}$. For isothermal processes, the left-hand side of the
equation is constant during the deflation, and we may choose the initial state
as the reference, where the shell volume is $V_0$ and the internal pressure
equal to some ambient pressure $p_a$, and so we have $F_\text{gas} = -p_a V_0
\ln V$. For the applied external pressure $p_\text{ex}$, an energy
contribution $p_\text{ex} V$ must be included. The total energy functional is
thus $G = U - p_a V_0 \ln V + p_\text{ex} V$, which is of the same form as
eq.\ (\ref{eq:G}). Note that in the undeformed configuration, the force balance
requires $p_\text{ex} = p_a$. The buckling of spherical shells with an
  internal ideal gas has in part been studied numerically in
  ref.\ \citenum{Pelekasis2014}.

\section{Bifurcation diagrams, stability discussion, and capsule collapse (snap-through)}

The shape equations are solved numerically as described in
ref.\ \citenum{Knoche2011} with the mechanical pressure $p$ as control
parameter. For the numerical analysis, it is convenient to choose a length
unit $R_0$ and tension unit $\EHo$. The shape equations then depend only on
the dimensionless pressure $p R_0/\EHo$, the Poisson ratio $\nu$ and the
dimensionless bending stiffness
\begin{equation}
\label{eq:FvK}
 \tilde E_B = \frac{E_B}{R_0^2 \, \EHo} 
 = \frac{H_0^2}{12 (1-\nu^2) R_0^2} = \frac{1}{\FvK}
\end{equation} 
which equals the inverse of the Föppl-von-K\'arm\'an-number $\gamma_\text{FvK}$.

Solutions for given elastic moduli and over a wide range of pressures $p$ have
been computed. From this dataset, bifurcation diagrams can be obtained
 from which
the stability in the three load cases -- mechanical pressure control, volume
control and osmotic pressure control -- can be derived. They contain different
solution branches,\cite{Knoche2011} and we concentrate here on the two most
relevant ones: One with uniformly contracted spherical shapes, and one with
buckled shapes with a single dimple. A third solution branch 
  are top-down symmetric shapes with two dimples; they have been shown to
  be less favorable for mechanical pressure control and volume control,
  \cite{Knoche2011} and it will be shown in section
  \ref{sec:enthalpy_landscape} that this is also true for osmotic pressure
  control.

\begin{figure*}[p]
 \centering
 \includegraphics[width=0.95\textwidth]{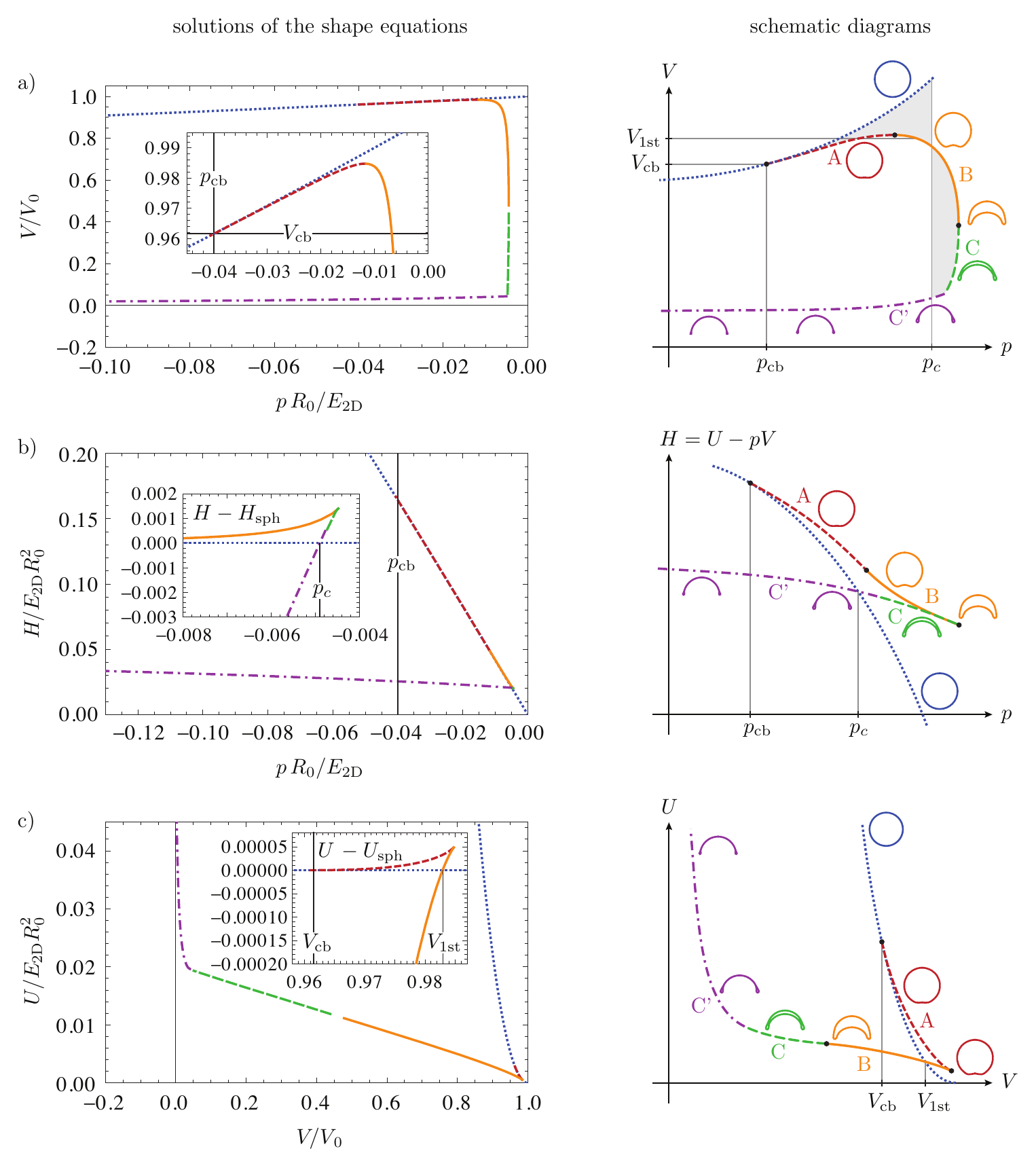}
 \caption[Bifurcation diagrams]
 {Bifurcation diagrams for buckling by mechanical pressure and volume
   control: a) volume-pressure relation, b) enthalpy as a function of the
 pressure, c) elastic energy as a function of the volume. The dotted blue
 line represents the spherical solution branch, the other colored lines
 represent buckled solution branches A, B, C and C' according to the labels
 and pictograms on the right. The insets in the energy diagrams in b) and c)
 show the differences between buckled and spherical branches. In all plots,
 the elastic moduli are $\tilde E_B = 10^{-4}$ and $\nu = 1/3$, and the same
 qualitative behavior has been obtained for all bending stiffnesses under
 consideration, from $\tilde E_B = 10^{-6}$ to $10^{-2}$, see also ref.\
 \citenum{Knoche2011}. On the right, schematic diagrams clarify the
 qualitative course of the solution branches.}
 \label{fig:bifdiag}
\end{figure*}

  The buckled branch with a single dimple develops from the spherical branch
  by flattening a region around one pole of the shell, then creating a dimple
  by inverting the region around the pole that subsequently grows until it
  finally leads to self-intersecting shapes. We suppress unphysical
  self-intersecting shapes by replacing them by shapes with opposite sides in
  contact, for which a simplified model has been developed in
  ref.\ \citenum{Knoche2011}. However, not all of the calculated
  shapes are stable. We
  split the buckled solution branch with a single dimple 
  into parts A, B, C and C', see
  fig.\ \ref{fig:bifdiag}, according to their stability under pressure and
  volume control as obtained from  bifurcation theory.
 Branches A, B, and C represent buckled shapes without opposite 
sides in contact, branch C' is the continuation of branch C after
  opposite sides made contact.

\subsection{Theorems from bifurcation theory}\label{sec:maddocks}

  We exploit very general mathematical theorems about
  the stability of the solution branches in bifurcation diagrams due to
  Maddocks \cite{Maddocks1987} in order to characterize
the stability of the different parts of the buckled solution branches. 
 A solution of the shape equations is only
  stable when it represents a local minimum of the energy functional (and not
  a maximum or saddle point).  The theorems from bifurcation theory allow us
  to infer stability from the slope of the volume-pressure relations and can
  be applied both to mechanical and osmotic pressure.

For the reader's convenience we will briefly summarize the relevant results of
ref.\ \citenum{Maddocks1987} concerning the stability of solution branches in a
bifurcation diagram. The solution branches shall originate from the
variational problem of minimizing a functional 
${\cal F}[\vec r, \lambda]$ with
respect to the function $\vec r$, while $\lambda$ is a bifurcation
parameter. 
For our buckling problems,
 ${\cal F}$ represents the total energy functional,
\emph{i.e.}, the above functionals $H$
for mechanical pressure control and $G$ for osmotic pressure control,
respectively,   and 
the bifurcation parameter $\lambda$ is the mechanical pressure 
$p$ or the external osmotic pressure $p_\text{ex}$, respectively.
The function $\vec r$ contains the parametrization of  the capsule shape. 
Specifically, we consider the case that the bifurcation parameter
enters the functional linearly in the form 
${\cal F}[\vec r, \lambda] = {\cal U}[\vec r] -
\lambda V[\vec r]$, which applies both to  mechanical and osmotic
pressure,  where  ${\cal U}$  is the corresponding energy 
of the shape $\vec r$ and $V$ its volume, \emph{cf.}\ eqs.\ (\ref{eq:delta_H}) and 
(\ref{eq:G}).

\begin{figure}
 \centering
 \includegraphics{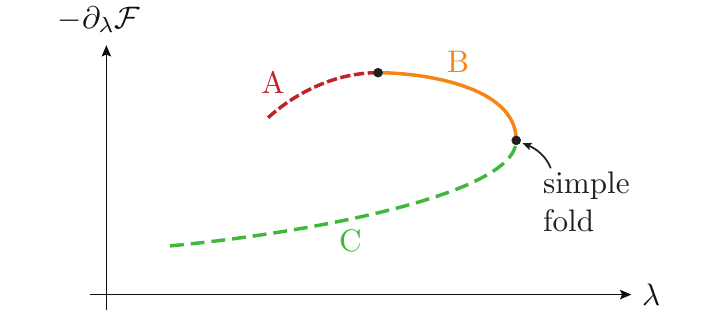}
 \caption{The distinguished bifurcation 
diagram with exemplary solution branches.
}
 \label{fig:maddocks}
\end{figure}

The solution branches $\vec r(\lambda)$ of this minimization problem are best
visualized in the \emph{distinguished bifurcation diagram} in which the
functional $-\partial_\lambda{{\cal F}}$, 
evaluated at a solution $\vec r(\lambda)$, is plotted
against the bifurcation parameter $\lambda$, see
fig.\ \ref{fig:maddocks}. Points of vertical tangency are called \emph{folds},
in our example this is the point between branches B and C.  

A solution branch is called \emph{stable} when it represents minima of the
functional ${\cal F}$. 
Mathematically, this is related to the second variation of
${\cal F}$: 
If it is positive definite in a solution $\vec r$, \emph{i.e.}, has only positive
eigenvalues, $\vec r$ is a minimum and, thus, a stable solution of the
minimization problem. We quote two results from ref.\ \citenum{Maddocks1987}
concerning the stability of solution branches: (i) The slope of a stable
solution branch in the distinguished bifurcation diagram is non-negative, and
(ii) the upper branch of a simple fold opening to the left has one more
negative eigenvalue than the lower branch. In the example of
fig.\ \ref{fig:maddocks}, A and C are candidates for stable branches according
to (i). However, (ii) states that the upper branch (consisting of A and B) of
the fold has one more negative eigenvalue than C. If C is stable, \emph{i.e.}, has no
negative eigenvalues, then A and B are unstable and have precisely one
negative eigenvalue.

Maddocks also discusses the  variational problem to minimize
the functional ${\cal U}[\vec r]$ under the constraint that $V[\vec r] = \rm
const$. He calls branches that are stable in this constrained problem
\emph{c-stable}. Stability in the constrained problem is a weaker condition
than stability in the unconstrained problem, because only variations that
leave $V$ constant can give rise to instabilities. 
Mathematically speaking, the second variation must be non-negative
on the tangent space to the constraint surface $V(\vec r) = \rm const$. 
\cite{Maddocks1987} Maddocks
shows that (iii) all stable branches are also c-stable, and (iv) the branches
that are c-stable but not stable are those with precisely one negative
eigenvalue and negative slope in the distinguished bifurcation diagram. In our
example of fig.\ \ref{fig:maddocks}, where we assume that A and B have one
negative eigenvalue, this means that branch B is c-stable.

The criteria (i) - (iv) can now be applied to study stability 
under  mechanical pressure control or  osmotic pressure 
control  and to study c-stability under volume control.

\subsection{Mechanical pressure control}

Let us start with the case of mechanical pressure control, which requires
unconstrained minimization of the enthalpy $H = U - pV$. This case has already
been discussed in ref.\ \citenum{Knoche2011}, we include it here for
completeness.  The bifurcation parameter is $p$, and the distinguished
  bifurcation diagram is the $V(p)$ diagram of fig.\ \ref{fig:bifdiag} a).
%
Branches A, B, and C/C' of the $V(p)$ diagram have the same structure as our
example in fig.\ \ref{fig:maddocks}. The $H(p)$ diagram in
fig.\ \ref{fig:bifdiag} b) reveals that the branches C and C' are stable: C'
seems to be the global enthalpy minimum over a large pressure range, and if
C' is stable, C must also be stable because the stability only changes at
folds. Thus, branch C/C' has only positive eigenvalues in the second
variation, and A and B have precisely one negative eigenvalue, and are
 therefore unstable under pressure control.

The bifurcation behavior under mechanical 
pressure control can thus be summarized as
 follows. When spherical shells are loaded with a negative internal pressure
 they remain spherical for small loads because the spherical branch is the
  global enthalpy minimum. At a critical pressure $p_c$, the branch C'
  (consisting of buckled shapes with self-contact) crosses the spherical
  branch in the $H(p)$ diagram.  Beyond this pressure, branch C' is the global
  energy minimum. Although it is energetically preferable for the shell to
  change from the spherical into a fully buckled shape at $p_c$, this will not
  happen spontaneously because both branches are metastable energy minima, and
 an energy barrier must be overcome. Spontaneous buckling is possible only at
  the classical buckling pressure\cite{Ventsel2001}
\begin{equation} \label{eq:pcb}
 p_\text{cb} = -4\sqrt{\tilde E_B} \EHo/R_0.
\end{equation}  
Here, the spherical solution branch becomes unstable, and the shell will
``fall'' from the spherical branch onto the branch C' where it is completely
collapsed (see pictograms in fig.\ \ref{fig:bifdiag} b) on the right).  This
direct transition into a completely collapsed state is also called
snap-through. Remarkably, the absolute value of $p_c$ is much smaller than
that of $p_\text{cb}$, for the elastic moduli of fig.\ \ref{fig:bifdiag}
approximately $p_c = 0.12 p_\text{cb}$ (below in eq.\ (\ref{eq:pc}), we will
give a more general analytical estimate for $p_c$). Our numerical studies show
that the complete collapse under pressure control happens on the whole
parameter range under investigation, $\tilde E_B = 10^{-6}$ to
$10^{-2}$. Although we cannot give a strict analytical argument we found no
numerical evidence that the qualitative behavior would change for even smaller
oder larger bending stiffnesses. 
We always find  that branch C'
rather than branch C crosses the spherical
  branch in the $H(p)$ diagram.
 This leads us to the conjecture that complete
capsule collapse is generic for buckling under pressure control.

\begin{figure*}
 \centering
 \includegraphics[width=\textwidth]{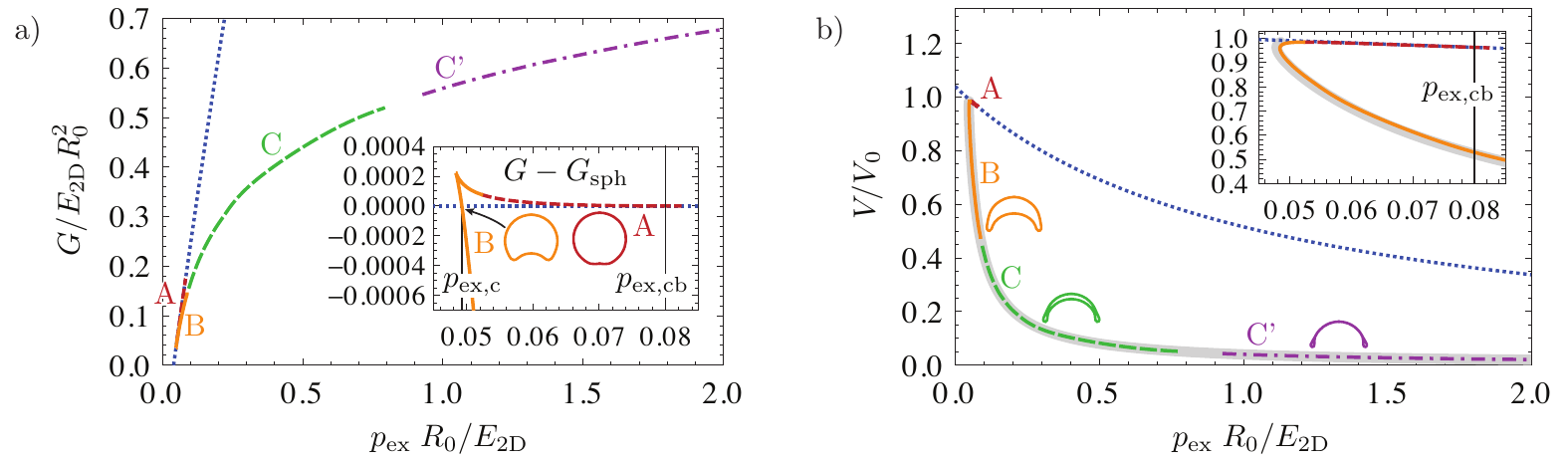}
 \caption[Bifurcation diagrams for osmotically induced buckling]
 {Bifurcation diagrams for osmotically induced buckling or buckling under
   pressure control with an internal gas. a) Energy $G$ as  a function of the
     external osmotic pressure. b) Volume-pressure relation. 
   The diagrams were created from
   the same data set used for fig.\ \ref{fig:bifdiag} (with $\tilde E_B =
   10^{-4}$, $\nu=1/3$ and $k_B T N_\text{in} = - p_{cb} V_0$), and the color
   code of the different branches is also the same. In comparison with
   fig.\ \ref{fig:bifdiag} a) and b) it should be noted that $p$ and
   $p_\text{ex}$ have different signs, and that a part of the orange branch B
   is stable now.  The thick gray line in the background of diagram b)
   represents the analytic result (\ref{eq:pex}) derived below.  }
 \label{fig:bifdiag_osmosis_gas}
\end{figure*}

\subsection{Volume control}

Stability under volume control corresponds to c-stability of shapes.  
Since we
have seen that branches A and B have precisely one negative eigenvalue of the
second variation and B has a negative slope $\partial_p{V} < 0$, we can
conclude that branch B is stable under volume control, but A is not. C and C'
are, of course, also stable under volume control. This is in accordance with
the $U(V)$ diagram, from which we see that branch B is the global energy
minimum when the volume is lowered beyond a critical volume $V_\text{1st}$. As
in the case of pressure control, buckling at this point involves overcoming an
energy barrier which can be read off from the inset in fig.\ \ref{fig:bifdiag}
c). This barrier vanishes at the classical buckling volume\cite{Quilliet2012}
\begin{equation} \label{eq:Vcb}
 \frac{V_\text{cb}}{V_0}
 \approx 1 - 6 (1-\nu) \sqrt{\tilde E_B} \qq{for} \tilde E_B \ll 1.
\end{equation} 
This behavior is analogous to the case of pressure control but, for volume
control, the first stable shapes after buckling are those of branch B, with a
medium large dimple, and not the completely collapsed ones of branch C' as for
pressure control.

Branch B, which contains the buckled shapes with small to medium sized dimples
that are frequently observed in microcapsule experiments,\cite{Fery2004,
  Quilliet2008, Sacanna2011, Okubo2003, Okubo2001} has thus a very interesting
property: It changes from stable to unstable when the mechanical pressure is
controlled instead of the volume. We will see that for osmotic pressure
control parts of branch B will become stabilized again.

Legendre transformations provide a link between the three bifurcation diagrams
in fig.\ \ref{fig:bifdiag}. The function $H(p)$ stems from the functional
$H[\vec r, p] = U[\vec r] - p V[\vec r]$ by inserting the numerical solutions
$\vec r(p)$ of the shape equations for a given pressure $p$, \emph{i.e.},
\begin{equation}
 H(p) = U[\vec r(p)] - p V[\vec r(p)].
\end{equation} 
Taking the derivative with respect to $p$, we must consider that the shape
changes by $\delta \vec r$ when the pressure is changed by $\diff p$. We thus
obtain
\begin{equation} \label{eq:dH/dp}
 \frac{\diff H}{\diff p} = \frac{\delta U - p \, \delta V}{\diff p} - V[\vec r(p)] = -V(p),
\end{equation} 
where we use $\delta U - p \, \delta V = 0$ because the shape equations were
derived from this condition. This result connects the $V(p)$ diagram, fig.\
\ref{fig:bifdiag} a), to the $H(p)$ diagram, fig.\ \ref{fig:bifdiag} b). Now, the function $U(V)$ is obtained as $U = H + pV$, or more precisely as
\begin{equation} \label{eq:U(V)}
 U(V) = H(p(V)) + p(V) V,
\end{equation} 
where $p(V)$ is the inverse function of $V(p)$. We recognize that the energy $U(V)$ is the \emph{Legendre transform} of the enthalpy $H(p)$, just like in thermodynamics\cite{Callen1960} from where our notation is adopted. Consequently, it follows that $\partial{U}/\partial{V} = p$ and that $H$ is also the Legendre transform of $U$.

The \emph{Maxwell construction}  from thermodynamics\cite{Callen1960,
  Knoche2011} can therefore be applied to the $V(p)$ diagram, in order to
construct the critical pressure $p_c$ and volume $V_\text{1st}$ of the
buckling transition. They are defined as the points in the energy diagrams
$H(p)$ and $U(V)$, respectively, where the buckled solution branch crosses the
spherical one. In the $V(p)$ diagram, the critical pressure
$p_c$ thus fulfills the condition of equal 
 shaded areas in fig.\ \ref{fig:bifdiag} a). 
The critical volume $V_\text{1st}$ can be constructed analogously,
with equal enclosed areas between the horizontal line $V_\text{1st}$ and the
spherical and buckled branches.

\subsection{Osmotic pressure control}

Let us now turn to the stability analysis for  osmotically induced buckling,
or buckling under pressure control with an internal gas. Now, the bifurcation parameter is the external part of
the osmotic pressure $p_\text{ex} = k_B T N_\text{ex} / V_\text{ex}$, because
this quantity can be controled in experiments by changing the concentration of
osmotically active particles outside the shell. In order to study stability
under osmotic pressure control we can use the available solutions of the shape
equations for mechanical pressure control, which have already been used to draw the bifurcation diagram fig.\ \ref{fig:bifdiag}. For each solution of the shape
equations for a given mechanical pressure $p$ and with a volume $V$, a
corresponding external osmotic pressure can be obtained as $p_\text{ex} =
p_\text{in} - p$ if a value for $k_B T N_\text{in}$ is chosen.

Figure~\ref{fig:bifdiag_osmosis_gas} shows the resulting bifurcation diagrams:
on the left, the energy diagram $G(p_\text{ex})$ and, on the right, the
reduced volume $V(p_\text{ex})/V_0$. The latter one is related to Maddock's
distinguished bifurcation diagram, since $-\partial_{p_\text{ex}} G = -V$, and
his stability discussion can be applied to the $V(p_\text{ex})$ diagram when
the minus sign is kept in mind. From both bifurcation diagrams it is evident
that, compared to pressure control without internal gas, some of the buckled
shapes of branch B are stabilized. To illustrate this, we use the same color
code for the  shapes as in fig.\ \ref{fig:bifdiag}, \emph{i.e.},
a shape corresponding to an orange point in fig.\ \ref{fig:bifdiag} also gives
an orange point in fig.\ \ref{fig:bifdiag_osmosis_gas}. Figure
\ref{fig:bifdiag_osmosis_gas} immediately shows that the buckled shape at the
critical external pressure is for osmotic pressure control a shape on branch
B, with a medium large dimple, rather than a collapsed state with opposite
sides in contact.

As for mechanical pressure control, there are also two critical external
osmotic pressures, $p_\text{ex,c}$ corresponding to the point where the
buckled and spherical branches cross in the energy diagram, and
$p_\text{ex,cb}$ corresponding to the classical buckling threshold, where 
the spherical shape becomes unstable and the
buckled branch separates from the spherical one. Again, the threshold
$p_\text{ex,c}$ where buckling becomes energetically favorable (but is only
accessible by overcoming an energy barrier, see the inset in
fig.\ \ref{fig:bifdiag_osmosis_gas}) is much smaller than the classical
threshold $p_\text{ex,cb}$ where the spherical branch loses its stability. The
latter value can be calculated as 
\begin{equation} \label{eq:pex,cb} 
p_\text{ex,cb} = {k_B T N_\text{in}}/{V_\text{cb}} - p_\text{cb}=
  p_\text{in}(V_\text{cb}) -  p_\text{cb}
 \end{equation} 
with $V_\text{cb}$ and $p_\text{cb}$ 
from eqs.\ (\ref{eq:Vcb}) and (\ref{eq:pcb}), respectively. 

How much of branch B becomes stabilized under osmotic pressure control 
and whether branch B (as in 
 the example shown in fig.\ \ref{fig:bifdiag_osmosis_gas}) or branch C or 
the snap-through branch C'  cross the spherical branch in the $G(p_\text{ex})$ 
  diagram depends
on the number of osmotically active particles  $N_\text{in}$
or the initial internal osmotic pressure: 
In the limit $N_\text{in} \rightarrow 0$, where there are no
osmotically active particles (or gas particles) enclosed in the shell, the
behavior for mechanical pressure control is recovered in which the whole
branch B is unstable and the first buckled shape after the instability is a
collapsed snap-through state on branch C'.
For an increasing number  $N_\text{in}$, we first find 
buckling into shapes C and, then, a stabilization of and buckling into 
branch B. Further increasing $N_\text{in}$
then  further extends 
the stabilized part of branch B. 
The bifurcation behavior under osmotic pressure control 
becomes 
 qualitatively similar to buckling under volume control if $N_\text{in}$
is sufficiently large such that
 the spherical branch exchanges stability with 
branch B as in the example shown in fig.\ \ref{fig:bifdiag_osmosis_gas}.

\section{Enthalpy landscape for buckled shapes: 
 osmotic pressure control and 
stabilization of non-collapsed  shapes}
\label{sec:enthalpy_landscape}

\begin{figure*}
 \centering
 \includegraphics{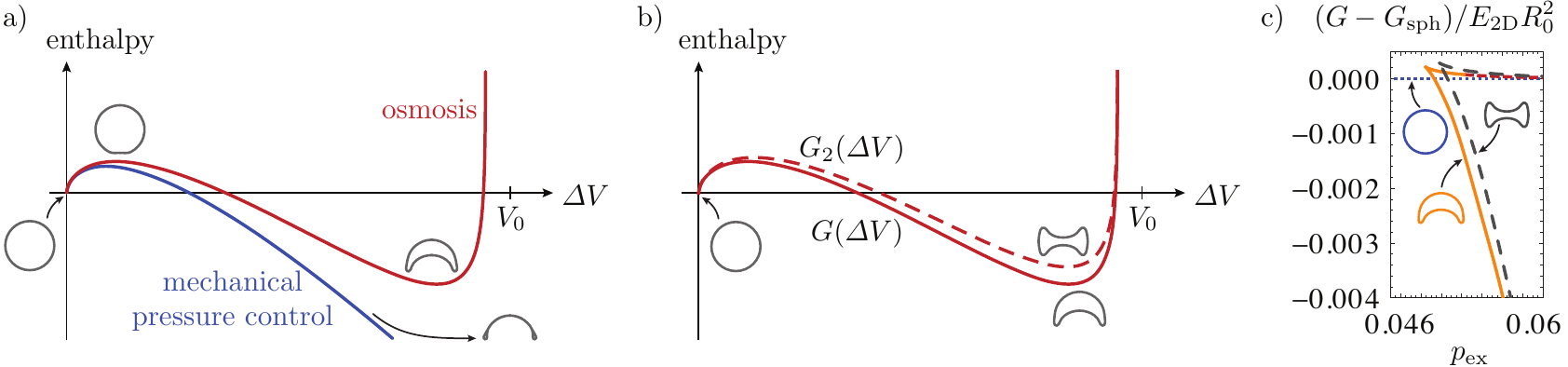}
 \caption[Enthalpy landscape for the buckling transition]
 {a) Enthalpy landscape for the buckling transition under mechanical pressure
   control (blue line) and osmotic pressure control (red line). After the
   enthalpy barrier has been overcome, mechanically pressurized shells can
   lower their enthalpy on and on by reducing the volume, but osmotically
   pressurized shells will end up in the enthalpy minimum at finite $\Delta
   V$. b) Effect of a second dimple in the shape (dashed line): The enthalpy
   function is raised, with the effect that the minimum of the function, where
   the stationary shape resides, is also lifted to higher enthalpy. c)
   Numerical demonstration (for the same parameters as in
   fig.\ \ref{fig:bifdiag_osmosis_gas}) that the branch with two dimples is
   energetically less favorable than the branch with one dimple. We plot
   the enthalpy difference to the spherical branch in order to better resolve
   the differences in the branches.}
 \label{fig:enthalpy_landscape}
\end{figure*}

The stabilizing effect of an internal medium on the non-collapsed shapes can
be shown more explicitly by considering the energy landscape during the
buckling process. The ``reaction coordinate'' that describes the progress of
buckling is $\Delta V = V_0 - V$. An analytic estimate of the elastic energy
in a shell with one dimple has been given by Pogorelov,\cite{Pogorelov1988}
\begin{equation}\label{eq:Upog}
 U_\text{Pog} \approx 2 \pi J_\text{min}
   \left( \frac{8}{3} \right)^{3/4} 
   \frac{\EHo}{\big( 1 - \nu^2 \big)^{1/4}} 
     \left( \tilde E_B \frac{\Delta V}{V_0} \right)^{3/4} R_0^2
\end{equation}
where $J_\text{min} = 1.15092$ is a numerical factor. For mechanical pressure
control, a term $- p V = -p (V_0 - \Delta V)$ must be added to obtain the
total energy (or enthalpy) $H(\Delta V) = U_\text{Pog}(\Delta V) + p \Delta V
+ \text{const}$. This results in a function $H(\Delta V) \sim \Delta V^{3/4} -
|p|\Delta V$ (because $p$ is negative) as plotted in fig.\ \ref{fig:enthalpy_landscape} a), blue line. There exists an energy barrier which has to be overcome, for example by manually indenting the shell, by imperfections or by thermal fluctuations, but once this is achieved, the shell tends to maximize  $\Delta V$ in order to minimize its energy. This means that, under pressure control, the  shell collapses completely upon buckling. This model is, of course, over-simplified because it relies on the Pogorelov model that becomes inaccurate for very large dimples.\cite{Pogorelov1988,Knoche2014,Knoche2014long} The shell cannot reach $\Delta V \geq V_0$, and even before there will be additional terms in the elastic energy caused by the constraint of no self-intersection.

The global minimum of $H(\Delta V)$ becomes  a boundary minimum at $\Delta V = V_0$ for pressure values  $|p|>|p_c|$. The criterion $H(0) = H(V_0$) thus provides an estimate for the critical pressure $p_c$, 
\begin{align} 
  p_c &= -2 \cdot 6^{1/4} J_\text{min} \, 
  \frac{\EHo \tilde E_B^{3/4}}{R_0 (1-\nu^2)^{1/4}}\,
\nonumber\\
   &= \frac{1}{2} \cdot 6^{1/4} J_\text{min}
      \frac{\tilde E_B^{1/4}}{ (1-\nu^2)^{1/4}}  p_\text{cb}
  \label{eq:pc}
 \end{align}
We checked with our numerical results the accuracy of this estimate over a large range of bending stiffnesses, from $\tilde E_B = 2 \cdot 10^{-6}$ to $10^{-3}$, and found that also the  numerical prefactor is in reasonable agreement with the numerical results, despite the  simplicity of the enthalpy landscape.

Pogorelov's model also becomes inaccurate for very small
dimples.\cite{Knoche2014,Knoche2014long} For the energy landscape, this has
the effect that the energy barrier is always present. The height of the energy
barrier is $H_\text{barrier} \sim {\EHo^4 \tilde E_B^3}/{|p|^3 R_0
  (1-\nu^2)}$. The barrier is even present for pressures $p$ exceeding the
critical buckling threshold $p_\text{cb}$, where buckling should become
spontaneous and a barrier should be absent. Therefore, one can simply assume
that small barriers $H_\text{barrier} \sim {\EHo^4 \tilde
  E_B^3}/{|p_\text{cb}|^3 R_0 (1-\nu^2)} \sim \EHo \tilde E_B^{3/2}R_0^2/
(1-\nu^2)$ can be overcome spontaneously. For an isotropic shell material,
with $\EHo = E H_0$ and $\tilde E_B = {EH_0^3}/({12(1-\nu^2)}$, this barrier
height corresponds to an indentation of the order of the shell thickness $H_0$
at the barrier. This argument is similar to a corresponding argument in 
ref.\ \citenum{Landau1970}, where  it is assumed that the
buckling threshold $p_\text{cb}$ can be identified with the necessary pressure
for an indentation of the order of the shell thickness $H_0$ to 
grow spontaneously. Apart from this problem for
pressures $p$ close to the buckling threshold $p_\text{cb}$, the energy
landscape is qualitatively correct for $|p| < |p_\text{cb}|$.

When we consider the appropriate energy functional for osmotic pressure or pressure control with an internal gas, a term $\propto - \ln V$ must be added to the energy functional. It penalizes small volumes and, therefore, prevents the shell volume from approaching  $\Delta V \rightarrow V_0$. The total energy (or free enthalpy) reads
\begin{equation} \label{eq:free_enthalpy_landscape}
 G(\Delta V) = U_\text{Pog}(\Delta V) - p_\text{ex} \Delta V 
    - k_B T N_\text{in} \ln(V_0 - \Delta V)
\end{equation} 
and has the qualitative shape  plotted in fig.\ \ref{fig:enthalpy_landscape} a), red line. There is no boundary minimum at $\Delta V = V_0$ corresponding to a fully collapsed state with $V=0$ but a local energy minimum at a finite volume, \emph{i.e.}, $\Delta V<V_0$. The volume at this minimum  depends on the elastic moduli, the external pressure $p_\text{ex}$ and the internal particle number $N_\text{in}$. This qualitatively explains why an internal gas or internal osmotically active particles prevent the full collapse of the shell and stabilize buckled shapes with medium volume reduction (parts of branch B).

It remains to justify why we concentrated our investigations on buckled shapes
with a single dimple only, and disregarded all other solution branches that
can be obtained from the shape equations.\cite{Knoche2011} 
Numerical solution of the shape equations in ref.\ \citenum{Knoche2011} 
have shown that all other solution branches are less favorable for 
volume control and mechanical pressure control. Here we present an 
analytical argument, which confirms these findings and also covers 
osmotic pressure control.
The most promising
candidates that could become energetically favorable for osmotic buckling are
shapes with multiple dimples. We can consider symmetric shapes with two
dimples within the Pogorelov model and within the axisymmetric shape equations
to show that their free enthalpy is larger than for one dimple. The volume
reduction $\Delta V$ of the shell is divided between the two dimples which
have $\Delta V/2$ each. According to the Pogorelov model, the elastic energy
of a double buckled shell is thus $U_\text{Pog 2}(\Delta V) = 2 \,
U_\text{Pog}(\Delta V/2) = 2^{1/4} \cdot U_\text{Pog}(\Delta V)$, where the
last equation holds because $U_\text{Pog} \sim \Delta V^{3/4}$. Thus, for
given volume difference it is energetically unfavorable to create multiple
dimples.\cite{Knoche2011,Quilliet2012}

Now we have to clarify how this translates to the free
enthalpy $G(p_\text{ex})$ for osmotic pressure control 
where a change of variables from $\Delta V$ to
$p_\text{ex}$ is necessary.
The branch with a single dimple has a free enthalpy
\begin{align}
G(p_\text{ex}) &= 
  \min_{\Delta V} \left[ U_\text{Pog}(\Delta V) - p_\text{ex} \Delta V
        - k_B T N_\text{in} \ln(V_0 - \Delta V) \right] \nonumber \\
         &\equiv \min_{\Delta V} \left[ f(\Delta V, \, p_\text{ex}) \right]
\end{align}
for osmosis. To obtain the enthalpy of the symmetrically buckled branch we just change $U_\text{Pog}$ to $U_\text{Pog 2}$ in this expression, which results in
\begin{equation}
G_2(p_\text{ex}) = \min_{\Delta V} 
 \left[ f(\Delta V, \, p_\text{ex}) + (2^{1/4}-1) U_\text{Pog}(\Delta V) \right]
\end{equation}
The additional term $(2^{1/4}-1) U_\text{Pog}(\Delta V)$ is positive for all
$\Delta V$. The volume-dependent enthalpy function whose minimum we are
searching is thus shifted to higher values, see
fig.\ \ref{fig:enthalpy_landscape} b). As a consequence, the stationary shape
that resides in the minimum is shifted to a higher enthalpy when there are two
dimples on the shell instead of one; and also the transition states at the
enthalpy maximum lie at higher enthalpy. This result is confirmed by the
enthalpy diagram fig.\ \ref{fig:enthalpy_landscape} c) that was generated from
the shape equations.

\section{Applications: shape control, 
shape analysis  and osmotic pressure sensing}

In osmotic buckling, both the external part $p_\text{ex} = k_B T
N_\text{ex}/V_\text{ex}$ of the osmotic pressure, which is given by the
external concentration of osmotically active particles, and the internal
particle number $N_\text{in}$, which is enclosed in the capsule during
synthesis, are relevant experimental control parameters. The external pressure
$p_\text{ex}$ allows to control the final buckled shape experimentally, and
the internal particle number $N_\text{in}$ allows to control the final buckled
shape and the buckling threshold $p_\text{ex,cb}$ itself. Both of these
controls provide interesting applications, which can be analyzed using the
energy landscape (\ref{eq:free_enthalpy_landscape}).

We can determine the energy minimum analytically and quantify the
concentration of osmotically active particles, which is needed inside and
outside the
shell in order to stabilize buckled shapes of a desired volume
reduction. Particularly interesting is the buckled shape that is obtained at
the buckling threshold (\ref{eq:pex,cb}), $p_\text{ex} = p_\text{ex,cb}$,
where the shell can buckle spontaneously. The condition for an extremum of the
free enthalpy is
\begin{equation}
 0 = G'(\Delta V) = U_\text{Pog}'(\Delta V) - p_\text{ex,cb} 
   + \frac{k_B T N_\text{in}}{V_0 - \Delta V}.
\label{eq:Gmin}
\end{equation} 
This equation can be solved for the internal osmolyte concentration
$N_\text{in}/V_0$ and simplifies considerably if only the leading order in
$\tilde E_B$ is retained. The value of $k_B T N_\text{in}$ also determines the
external pressure $p_\text{ex,cb}$ needed to induce buckling, see
eq.\ (\ref{eq:pex,cb}). For both values, the simplified results are
\begin{equation}\label{eq:Nin_and_pex}
 \begin{aligned} 
 k_B T \frac{N_\text{in}}{V_0} &\approx 
  4 \left(\frac{V_0}{\Delta V} - 1\right) 
     \frac{\EHo}{R_0} \sqrt{\tilde E_B} \\ \text{and} \qquad 
 p_\text{ex,cb} &\approx 4 \frac{V_0}{\Delta V} 
    \frac{\EHo}{R_0} \sqrt{\tilde E_B}.
\end{aligned} 
\end{equation} 
Both results can be directly translated into concentrations of osmotically
active particles inside and outside the shell. The classical buckling pressure
$p_\text{cb} = -4 \sqrt{\tilde E_B} \EHo/R_0$ occurs as the relevant scale in
eq.\ (\ref{eq:Nin_and_pex}). In order to obtain buckled shapes with $\Delta V =
V_0/2$, for example, one should adjust the internal osmolyte concentration to
${N_\text{in}}/{V_0} = -p_\text{cb}/k_BT$  and the external osmolyte
concentration to ${N_\text{ex}}/{V_\text{ex}} = -2p_\text{cb}/k_BT$. These are
exactly the values used in fig.\ \ref{fig:bifdiag_osmosis_gas}, and the inset
in the $V(p_\text{ex})$ diagram confirms that the buckling at the classical
threshold indeed results in a shape close to $V=V_0/2$.

Because the external osmotic pressure determines the volume of the buckled
capsule, we can also use the shape or volume of osmotically buckled capsules
as an indicator for the applied osmotic pressure. Solving the equation $G'(\Delta V) = 0$, for  $p_\text{ex}$ we
find the relation between capsule volume and external osmotic pressure
\begin{align} 
p_\text{ex} &= p_\text{in,0} \left(1- \frac{\Delta V}{V_0} \right)^{-1}
  \nonumber\\
  &~+ \frac{3}{2} 6^{1/4} J_\text{min} \, 
  \frac{\EHo \tilde E_B^{3/4}}{R_0 (1-\nu^2)^{1/4}}
     \left( \frac{\Delta V}{V_0} \right)^{-1/4}
\label{eq:pex}
\end{align} 
with the internal osmotic  pressure in 
the undeformed reference state, $p_\text{in,0} = k_B T N_\text{in} /V_0$.
This relation has a simple interpretation:
The first term in eq.\ (\ref{eq:pex}) would be the 
relation between external osmotic pressure and capsule volume
 if the capsule exactly assumed its
osmotically preferred volume $V = p_\text{in,0} V_0/p_\text{ex}$.
The second term captures the additional influence of 
  shell elasticity on this relation.

The relation (\ref{eq:pex}) 
matches the numerical results with a striking accuracy as can be
seen in the bifurcation diagram fig.\ \ref{fig:bifdiag_osmosis_gas} (gray
line). Because the free enthalpy landscape is based on the approximate
Pogorelov model, which is inaccurate for large dimples, we would expect our
analytic estimate also to become inaccurate for large $\Delta
V$. Surprisingly, this is not the case. For large $\Delta V$, the position of
the free enthalpy minimum is primarily determined by the competition of the osmotic
terms $- p_\text{ex} \Delta V$ and $- k_B T N_\text{in} \ln(V_0 - \Delta V)$
in eq.\ (\ref{eq:free_enthalpy_landscape}); 
the elastic energy $U_\text{Pog}$ plays
a subordinate role. Indeed, the purely osmotic
approximation $p_\text{ex} = k_B T N_\text{in}
/ (V_0 - \Delta V)$, where the elastic contribution is completely neglected,
is in good agreement with the numerical pressure-volume-relation for $\Delta V
\gtrsim 0.5$. Neglecting the elastic contribution in eq.\ (\ref{eq:pex}) is
justified for small $\tilde E_B$ (and not too small $\Delta V$) because $k_B T
N_\text{in} = \order{p_\text{cb}} \sim \tilde E_B^{1/2}$ and the elastic term
is $\sim \tilde E_B^{3/4}$.

\begin{figure*}
 \centering
 \includegraphics{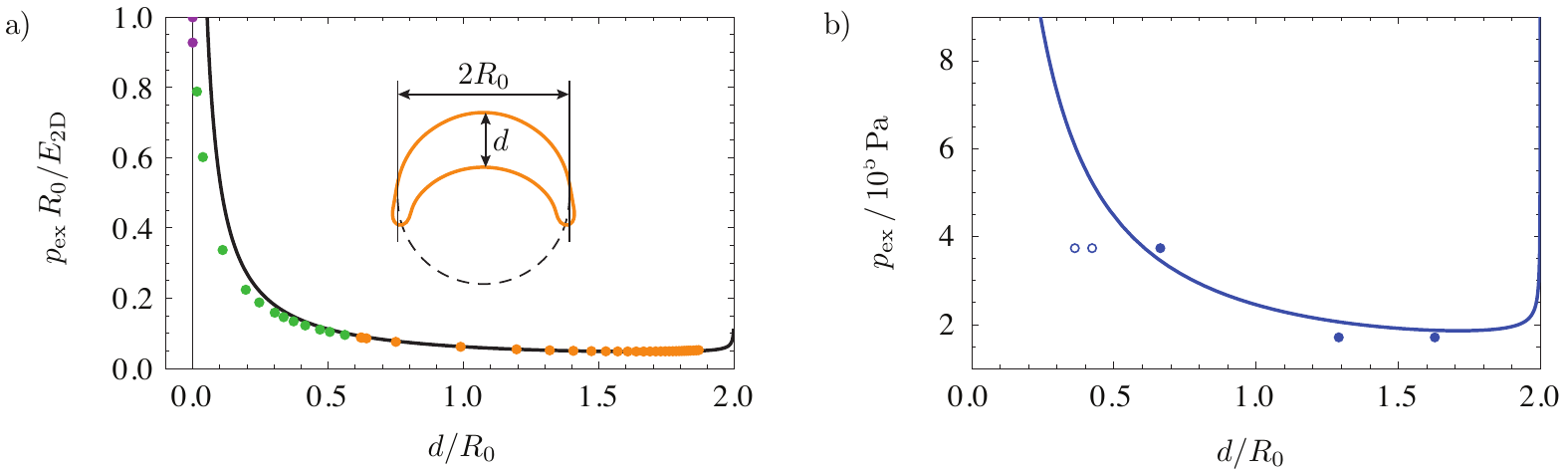}
 \caption{ Using a buckled shell as an osmotic pressure sensor: From a
   measurement of the depth $d$ and original shell radius $R_0$, the external
   osmotic pressure $p_\text{ex}$ can be determined. a) The data points are
   generated from the data set already used in figs.\ \ref{fig:bifdiag} and
   \ref{fig:bifdiag_osmosis_gas}, with $\tilde E_B = 10^{-4}$, $\nu=1/3$ and
   $k_B T N_\text{in} = - p_{cb} V_0$, and the solid line represents the
   analytic approximation based on eq.\ (\ref{eq:pex}).  b) Analysis of
   experimental results published in ref.\ \citenum{Gao2001}. The data points
   from the experiments are fitted using eq.\ (\ref{eq:pex}) with the internal
   osmotic pressure as fitting parameter. The open points were excluded from
   the fit because the experimental images looked conspicious that they may
   not represent centered cross-sections.  }
\label{fig:osm_pressure_sensor}
\end{figure*}

Equation (\ref{eq:pex}) provides the basis for measurements of the external
osmotic pressure by using elastic capsules as pressure sensors. The capsules
must be ``calibrated'' in the sense that their elastic properties, size and
internal osmolyte concentration are known. When they are embedded in a bath
with a larger, unknown osmolyte concentration and buckle consequently, their
volume difference can be measured and inserted into eq.\ (\ref{eq:pex}) to obtain
$p_\text{ex}$ or the external osmolyte concentration $N_\text{ex}/V_\text{ex}
= p_\text{ex}/k_B T$. The volume measurement could be achieved through a
microscopy image analysis, in the simplest version by measuring the shell
depth $d$ and original radius $R_0$ (see fig.\ \ref{fig:osm_pressure_sensor})
and using the geometrical relation for shapes whose dimple is an exact
mirror-reflection of a spherical cap\cite{Knoche2014long} to obtain $\Delta
V/V_0 = (1-d/2R_0)^2 (2+d/2R_0)/2$. While the relation (\ref{eq:pex}) for
$p_\text{ex}(\Delta V)$ is very precise, this relation $\Delta V(d)$ acquires
some errors, but fig.\ \ref{fig:osm_pressure_sensor} a) shows that these
errors are only significant for $d \lesssim R_0/2$.

Vice versa, 
eq.\ (\ref{eq:pex}) or the resulting relation for 
$p_\text{ex}$  as a function of $d/R_0$, see
fig.\ \ref{fig:osm_pressure_sensor}, 
can be used to determine the capsule's material parameters 
by fitting experimental data
for  $d/R_0$ at different external osmotic pressures 
$p_\text{ex}$.
Specifically, eq.\ (\ref{eq:pex}) can be used to determine the 
parameter combination  $\EHo\tilde E_B^{3/4}/R_0$
 and the internal osmotic pressure  $p_\text{in,0}$. 
In combination with an  analysis  of the 
maximal edge curvature of buckled shapes as proposed in ref.\
\citenum{Knoche2011} and experimentally realized in ref.\ \citenum{Jose2014},
which allows to determine the reduced bending modulus $\tilde E_B$, 
both  elastic moduli and the internal osmotic pressure can be obtained 
from relatively simple shape analyses of osmotically pressurized shells. To this end, accurate measurements of the external osmotic pressure and images of cross-sections along the axis of symmetry of the shells must be provided.

We tested such an analysis using the data published in
ref.\ \citenum{Gao2001} for polyelectrolyte capsules with radius $R_0 = 2\cdot
10^{-6} \, \rm m$ and wall thickness $H_0 = 2 \cdot 10^{-8} \, \rm m$.  The
polyelectrolyte capsules were then deflated osmotically, by adding
poly(styrene sulfonate, sodium salt) (PSS) to the exterior solution. The
osmotically active particles are the counter-ions surrounding the PSS
molecules, and they exert an external osmotic pressure $p_\text{ex}$ on the
capsules. In the experiments, the values of $p_\text{ex}$ were measured with a
Vapor Pressure Osmometer.
In view of the few available data points, which can be obtained 
from the confocal microscopy 
capsule images  in  ref.\ \citenum{Gao2001}, 
we use the value for the shear modulus of the shell 
material  $G =500 \, \rm MPa$ given in ref.\ \citenum{Gao2001}, which
corresponds to a Young modulus of $E = 1500 \, \rm MPa$ if $\nu =
0.5$.
Using also the measured values for capsule radius and thickness 
 this leads to $\EHo=30\,{\rm N/m}$ and 
 $\tilde E_B = 1.11 \cdot 10^{-5}$. 
Inside the capsule we also expect a certain concentration of ions, because the
capsule was fabricated from polyelectrolytes. This gives rise 
to a nonzero but unknown internal osmotic pressure
$p_\text{in,0} = k_B T N_\text{in}/V_0$ (in the undeformed state) 
which serves as
the only  fitting parameter in the following in order to
explain the observed shapes after osmotic buckling.

The value for $G$  obtained 
 in  ref.\ \citenum{Gao2001} might be questionable because
its determination relied on a measurement of the buckling pressure
using the classical buckling pressure 
$|p_\text{cb}|$, see eq.\ (\ref{eq:pcb}).
This determination assumes a vanishing internal 
pressure, \emph{i.e.}, $p_\text{in} \approx 0$ in eq.\ (\ref{eq:pex,cb}) and, 
moreover, the classical buckling pressure (\ref{eq:pex,cb}) only
 represents  an upper bound for the 
buckling pressure. Real imperfect shells buckle already at considerably
weaker pressures, \cite{Carlson1967,Hutchinson1967} between
the classical osmotic buckling pressure $p_\text{ex,cb}$,  where 
the spherical shape becomes unstable and the 
much smaller critical osmotic pressure  $p_\text{ex,c}$, where buckling 
becomes energetically favorable as discussed above.
As already pointed out, values for  $\EHo$ and $\tilde E_B$ 
could  also be obtained 
from a shape analysis, in principle,  if shape images for 
more external osmotic pressures 
$p_\text{ex}$ were available.

From five confocal microscopy images, figs.\ 2 (b) and (c) in
ref.\ \citenum{Gao2001}, we measured the ratio $d/R_0$. An uncertainty 
arises because we are not sure if the cross-sections imaged
by the confocal microscopy cut through the center of the capsules and if they
are oriented along the axis of symmetry of the capsules. For each 
image, the external osmotic pressure was given in ref.\ \citenum{Gao2001}. The
resulting data points are plotted in fig.\ \ref{fig:osm_pressure_sensor} b),
together with the fit using eq.\ (\ref{eq:pex}). 
For the fit parameter we obtained
$k_B T N_\text{in} = 5.4 \cdot 10^{-12} \, \rm J$, which corresponds to an
internal osmotic 
pressure (in the undeformed state) of 
$p_\text{in,0} = k_B T N_\text{in} /V_0 = 1.6
\cdot 10^5 \, \rm Pa$ and to a concentration of $N_\text{in} /V_0 = 65 \, \rm
mol/m^3$. Equation (\ref{eq:pex}) describes the experimental results
with reasonable accuracy.

\section{Discussion and Conclusion}

We have shown that the stability of buckled spherical shells (with respect to
axisymmetric deformation modes) depends on the specific system that generates
the pressure difference between the inside and outside. If a simple mechanical
pressure difference is prescribed, the enclosed volume will not affect the
applied mechanical pressure, and the shell will collapse completely after the
buckling has set in. This is known as snap-through buckling in the shell
theory literature. On the other hand, when the system is constructed so that
the shell must have a given volume, the first stable shapes after buckling
have a small, but finite dimple.

In most experiments, there will be a feedback between the deformation and the
pressure difference exerted on the shell, for example, for osmotic buckling or
if the shell encloses a gas. The feedback by an internal medium will stop the
snap-through buckling at a finite volume, thus stabilizing buckled shapes with
medium volume. Our findings explain why these are the shapes that are usually
observed in experiments, although they are unstable from the simple viewpoint
of pressure control.

The stabilizing effect of an internal medium is quite generic as long as the
force density exerted on the shell is still a normal pressure that is
spatially constant. We checked that the same qualitative results could be
obtained by including a compressible fluid in the shell, with an energy
contribution $F \propto (V-V_0)^2$. The reason for this generic behavior is
that the enthalpy landscape, see fig.\ \ref{fig:enthalpy_landscape}, is qualitatively identical, no matter how exactly the energies that penalize large volume differences look like.

Within this paper we specifically discussed buckling under (i) volume control,
(ii) mechanical pressure control and (iii) osmotic pressure control. Yet, even
more experimental situations are conceivable, which give rise to a feedback
between volume and pressure difference: (iv) As already mentioned, the shell
can be filled with a {\em compressible} fluid.  (v) The elastic properties of
the shell could depend on the concentration of an enclosed substance,
 \emph{e.g.}, if
the substance chemically reacts with the shell material. This will give rise
to capsule volume dependent elastic properties. (vi) One frequently used
mechanism in volume controled experiments is to slowly dissolve the interior
liquid of the capsule by the external liquid, thus reducing the internal
volume. \cite{Zoldesi2008,Quilliet2008, Datta2010} This procedure will involve
feedback as soon as the exterior volume is no longer much larger than the
internal capsule volume. If reducing the capsule volume increases the internal
pressure or stiffens the capsule material, such feedback mechanisms will
stabilize non-collapsed buckled shapes.  If a reduced capsule volume increases
the external pressure or softens the capsule material, complete collapse upon
buckling will be the generic behavior.

For osmotic pressure control, the capsule tends to assume a preferred
volume which is prescribed by 
the osmolyte concentrations. 
Therefore, the observed shape bifurcation behavior for osmotic 
pressure control becomes  typically qualititatively similar to
   buckling under volume control, see figs.\ \ref{fig:bifdiag}
  and \ref{fig:bifdiag_osmosis_gas}. In particular, 
snap-through buckling is suppressed. 
  This requires, however, 
  that  the initial
  osmolyte concentration in the capsule interior is sufficiently 
  large.  We presented a
   quantitative theory which also captures the influence of 
  shell elasticity on the resulting relation (\ref{eq:pex}) 
  between external osmotic pressure and  capsule volume.
 Buckling under osmotic pressure is indeed intermediate 
 between buckling under volume control and buckling under mechanical pressure:
In the limit of a small number $N_\text{in}$
  of osmotically active molecules in the capsule interior,
  buckling under mechanical pressure control is recovered;
 for  increasing  $N_\text{in}$, the behavior effectively approaches 
  buckling under volume control.

We have shown that these findings can be 
 relevant for the control of buckled shapes in
  applications by controling  the osmolyte concentration.  
  Conversely, we can use elastic capsules 
  as  osmotic pressure sensors, and an accurate analytic formula is derived that
  allows to deduce the  osmotic pressure from the observed volume of buckled
  capsules using eq.\ (\ref{eq:pex}). 
 This relation can also be used to obtain elastic moduli of the capsule 
 and its internal osmotic pressure from shape changes of the capsule 
  if the external osmotic pressure is experimentally controled. 
We applied this procedure to published 
  experimental data from Gao {\it et al.} \cite{Gao2001} 
on polyelectrolyte capsules. 
  Our findings are also relevant for stabilizing 
   buckled shapes of a desired volume in
  applications by choosing the osmolyte concentrations according to
  eq.\ (\ref{eq:Nin_and_pex}) to realize a desired $\Delta V$.

\footnotesize{
\bibliography{Literatur} 

\providecommand*{\mcitethebibliography}{\thebibliography}
\csname @ifundefined\endcsname{endmcitethebibliography}
{\let\endmcitethebibliography\endthebibliography}{}
\begin{mcitethebibliography}{39}
\providecommand*{\natexlab}[1]{#1}
\providecommand*{\mciteSetBstSublistMode}[1]{}
\providecommand*{\mciteSetBstMaxWidthForm}[2]{}
\providecommand*{\mciteBstWouldAddEndPuncttrue}
  {\def\EndOfBibitem{\unskip.}}
\providecommand*{\mciteBstWouldAddEndPunctfalse}
  {\let\EndOfBibitem\relax}
\providecommand*{\mciteSetBstMidEndSepPunct}[3]{}
\providecommand*{\mciteSetBstSublistLabelBeginEnd}[3]{}
\providecommand*{\EndOfBibitem}{}
\mciteSetBstSublistMode{f}
\mciteSetBstMaxWidthForm{subitem}
{(\emph{\alph{mcitesubitemcount}})}
\mciteSetBstSublistLabelBeginEnd{\mcitemaxwidthsubitemform\space}
{\relax}{\relax}

\bibitem[Discher \emph{et~al.}(1998)Discher, Boal, and Boey]{Discher1998}
D.~E. Discher, D.~H. Boal and S.~K. Boey, \emph{Biophys. J.}, 1998,
  \textbf{75}, 1584--97\relax
\mciteBstWouldAddEndPuncttrue
\mciteSetBstMidEndSepPunct{\mcitedefaultmidpunct}
{\mcitedefaultendpunct}{\mcitedefaultseppunct}\relax
\EndOfBibitem
\bibitem[Michel \emph{et~al.}(2006)Michel, Ivanovska, Gibbons, Klug, Knobler,
  Wuite, and Schmidt]{Michel2006}
J.~P. Michel, I.~L. Ivanovska, M.~M. Gibbons, W.~S. Klug, C.~M. Knobler,
  G.~J.~L. Wuite and C.~F. Schmidt, \emph{Proc. Natl. Acad. Sci. U. S. A.},
  2006, \textbf{103}, 6184--9\relax
\mciteBstWouldAddEndPuncttrue
\mciteSetBstMidEndSepPunct{\mcitedefaultmidpunct}
{\mcitedefaultendpunct}{\mcitedefaultseppunct}\relax
\EndOfBibitem
\bibitem[Katifori \emph{et~al.}(2010)Katifori, Alben, Cerda, Nelson, and
  Dumais]{Katifori2010}
E.~Katifori, S.~Alben, E.~Cerda, D.~R. Nelson and J.~Dumais, \emph{Proc. Natl.
  Acad. Sci. U. S. A.}, 2010, \textbf{107}, 7635--9\relax
\mciteBstWouldAddEndPuncttrue
\mciteSetBstMidEndSepPunct{\mcitedefaultmidpunct}
{\mcitedefaultendpunct}{\mcitedefaultseppunct}\relax
\EndOfBibitem
\bibitem[Meier(2000)]{Meier2000}
W.~Meier, \emph{Chem. Soc. Rev.}, 2000, \textbf{29}, 295\relax
\mciteBstWouldAddEndPuncttrue
\mciteSetBstMidEndSepPunct{\mcitedefaultmidpunct}
{\mcitedefaultendpunct}{\mcitedefaultseppunct}\relax
\EndOfBibitem
\bibitem[Fery \emph{et~al.}(2004)Fery, Dubreuil, and M\"{o}hwald]{Fery2004}
A.~Fery, F.~Dubreuil and H.~M\"{o}hwald, \emph{New J. Phys.}, 2004, \textbf{6},
  18--18\relax
\mciteBstWouldAddEndPuncttrue
\mciteSetBstMidEndSepPunct{\mcitedefaultmidpunct}
{\mcitedefaultendpunct}{\mcitedefaultseppunct}\relax
\EndOfBibitem
\bibitem[Neubauer \emph{et~al.}(2013)Neubauer, Poehlmann, and
  Fery]{Neubauer2013}
M.~P. Neubauer, M.~Poehlmann and A.~Fery, \emph{Adv. Colloid Interface Sci.},
  2013, \textbf{207}, 65--80\relax
\mciteBstWouldAddEndPuncttrue
\mciteSetBstMidEndSepPunct{\mcitedefaultmidpunct}
{\mcitedefaultendpunct}{\mcitedefaultseppunct}\relax
\EndOfBibitem
\bibitem[Rehage \emph{et~al.}(2002)Rehage, Husmann, and Walter]{Rehage2002}
H.~Rehage, M.~Husmann and A.~Walter, \emph{Rheol. Acta}, 2002, \textbf{41},
  292\relax
\mciteBstWouldAddEndPuncttrue
\mciteSetBstMidEndSepPunct{\mcitedefaultmidpunct}
{\mcitedefaultendpunct}{\mcitedefaultseppunct}\relax
\EndOfBibitem
\bibitem[Donath \emph{et~al.}(1998)Donath, Sukhorukov, Caruso, Davis, and
  M\"{o}hwald]{Donath1998}
E.~Donath, G.~Sukhorukov, F.~Caruso, S.~Davis and H.~M\"{o}hwald, \emph{Ang.
  Chem. Int. Ed.}, 1998, \textbf{37}, 2201\relax
\mciteBstWouldAddEndPuncttrue
\mciteSetBstMidEndSepPunct{\mcitedefaultmidpunct}
{\mcitedefaultendpunct}{\mcitedefaultseppunct}\relax
\EndOfBibitem
\bibitem[Gao \emph{et~al.}(2001)Gao, Donath, Moya, Dudnik, and
  M\"{o}hwald]{Gao2001}
C.~Gao, E.~Donath, S.~Moya, V.~Dudnik and H.~M\"{o}hwald, \emph{Eur. Phys. J.
  E}, 2001, \textbf{5}, 21--27\relax
\mciteBstWouldAddEndPuncttrue
\mciteSetBstMidEndSepPunct{\mcitedefaultmidpunct}
{\mcitedefaultendpunct}{\mcitedefaultseppunct}\relax
\EndOfBibitem
\bibitem[Okubo \emph{et~al.}(2001)Okubo, Minami, and Morikawa]{Okubo2001}
M.~Okubo, H.~Minami and K.~Morikawa, \emph{Colloid Polym. Sci.}, 2001,
  \textbf{279}, 931--935\relax
\mciteBstWouldAddEndPuncttrue
\mciteSetBstMidEndSepPunct{\mcitedefaultmidpunct}
{\mcitedefaultendpunct}{\mcitedefaultseppunct}\relax
\EndOfBibitem
\bibitem[Okubo \emph{et~al.}(2003)Okubo, Minami, and Morikawa]{Okubo2003}
M.~Okubo, H.~Minami and K.~Morikawa, \emph{Colloid Polym. Sci.}, 2003,
  \textbf{281}, 214--219\relax
\mciteBstWouldAddEndPuncttrue
\mciteSetBstMidEndSepPunct{\mcitedefaultmidpunct}
{\mcitedefaultendpunct}{\mcitedefaultseppunct}\relax
\EndOfBibitem
\bibitem[Zoldesi \emph{et~al.}(2008)Zoldesi, Ivanovska, Quilliet, Wuite, and
  Imhof]{Zoldesi2008}
C.~I. Zoldesi, I.~L. Ivanovska, C.~Quilliet, G.~J.~L. Wuite and A.~Imhof,
  \emph{Phys. Rev. E}, 2008, \textbf{78}, 1--8\relax
\mciteBstWouldAddEndPuncttrue
\mciteSetBstMidEndSepPunct{\mcitedefaultmidpunct}
{\mcitedefaultendpunct}{\mcitedefaultseppunct}\relax
\EndOfBibitem
\bibitem[Quilliet \emph{et~al.}(2008)Quilliet, Zoldesi, Riera, van Blaaderen,
  and Imhof]{Quilliet2008}
C.~Quilliet, C.~Zoldesi, C.~Riera, A.~van Blaaderen and A.~Imhof, \emph{Eur.
  Phys. J. E}, 2008, \textbf{27}, 13--20\relax
\mciteBstWouldAddEndPuncttrue
\mciteSetBstMidEndSepPunct{\mcitedefaultmidpunct}
{\mcitedefaultendpunct}{\mcitedefaultseppunct}\relax
\EndOfBibitem
\bibitem[Sacanna \emph{et~al.}(2011)Sacanna, Irvine, Rossi, and
  Pine]{Sacanna2011}
S.~Sacanna, W.~T.~M. Irvine, L.~Rossi and D.~J. Pine, \emph{Soft Matter}, 2011,
  \textbf{7}, 1631\relax
\mciteBstWouldAddEndPuncttrue
\mciteSetBstMidEndSepPunct{\mcitedefaultmidpunct}
{\mcitedefaultendpunct}{\mcitedefaultseppunct}\relax
\EndOfBibitem
\bibitem[Datta \emph{et~al.}(2012)Datta, Kim, Paulose, Abbaspourrad, Nelson,
  and Weitz]{Datta2012}
S.~S. Datta, S.-H. Kim, J.~Paulose, A.~Abbaspourrad, D.~Nelson and D.~Weitz,
  \emph{Phys. Rev. Lett.}, 2012, \textbf{109}, 1--5\relax
\mciteBstWouldAddEndPuncttrue
\mciteSetBstMidEndSepPunct{\mcitedefaultmidpunct}
{\mcitedefaultendpunct}{\mcitedefaultseppunct}\relax
\EndOfBibitem
\bibitem[Knoche and Kierfeld(2011)]{Knoche2011}
S.~Knoche and J.~Kierfeld, \emph{Phys. Rev. E}, 2011, \textbf{84}, 046608\relax
\mciteBstWouldAddEndPuncttrue
\mciteSetBstMidEndSepPunct{\mcitedefaultmidpunct}
{\mcitedefaultendpunct}{\mcitedefaultseppunct}\relax
\EndOfBibitem
\bibitem[Jose \emph{et~al.}(2014)Jose, Kamp, van Blaaderen, and
  Imhof]{Jose2014}
J.~Jose, M.~Kamp, A.~van Blaaderen and A.~Imhof, \emph{Langmuir}, 2014,
  \textbf{30}, 2385--93\relax
\mciteBstWouldAddEndPuncttrue
\mciteSetBstMidEndSepPunct{\mcitedefaultmidpunct}
{\mcitedefaultendpunct}{\mcitedefaultseppunct}\relax
\EndOfBibitem
\bibitem[Knoche \emph{et~al.}(2013)Knoche, Vella, Aumaitre, Degen, Rehage,
  Cicuta, and Kierfeld]{Knoche2013}
S.~Knoche, D.~Vella, E.~Aumaitre, P.~Degen, H.~Rehage, P.~Cicuta and
  J.~Kierfeld, \emph{Langmuir}, 2013, \textbf{29}, 12463--71\relax
\mciteBstWouldAddEndPuncttrue
\mciteSetBstMidEndSepPunct{\mcitedefaultmidpunct}
{\mcitedefaultendpunct}{\mcitedefaultseppunct}\relax
\EndOfBibitem
\bibitem[Ventsel and Krauthammer(2001)]{Ventsel2001}
E.~Ventsel and T.~Krauthammer, \emph{{Thin Plates and Shells}}, CRC Press,
  2001\relax
\mciteBstWouldAddEndPuncttrue
\mciteSetBstMidEndSepPunct{\mcitedefaultmidpunct}
{\mcitedefaultendpunct}{\mcitedefaultseppunct}\relax
\EndOfBibitem
\bibitem[Landau and Lifshitz(1970)]{Landau1970}
L.~Landau and E.~Lifshitz, \emph{{Theory of Elasticity}}, Pergamon Press,
  Oxford, 2nd edn., 1970\relax
\mciteBstWouldAddEndPuncttrue
\mciteSetBstMidEndSepPunct{\mcitedefaultmidpunct}
{\mcitedefaultendpunct}{\mcitedefaultseppunct}\relax
\EndOfBibitem
\bibitem[Hutchinson(1967)]{Hutchinson1967}
J.~W. Hutchinson, \emph{J. Appl. Mech.}, 1967, \textbf{34}, 49\relax
\mciteBstWouldAddEndPuncttrue
\mciteSetBstMidEndSepPunct{\mcitedefaultmidpunct}
{\mcitedefaultendpunct}{\mcitedefaultseppunct}\relax
\EndOfBibitem
\bibitem[Koiter(1969)]{Koiter1969}
W.~Koiter, \emph{Proc. Kon. Nederl. Akad. Wet. Amsterdam B}, 1969, \textbf{72},
  40\relax
\mciteBstWouldAddEndPuncttrue
\mciteSetBstMidEndSepPunct{\mcitedefaultmidpunct}
{\mcitedefaultendpunct}{\mcitedefaultseppunct}\relax
\EndOfBibitem
\bibitem[Timoshenko and Woinowsky-Krieger(1959)]{Timoshenko1959}
S.~P. Timoshenko and S.~Woinowsky-Krieger, \emph{{Theory of Plates and
  Shells}}, McGraw-Hill, New York, 2nd edn., 1959\relax
\mciteBstWouldAddEndPuncttrue
\mciteSetBstMidEndSepPunct{\mcitedefaultmidpunct}
{\mcitedefaultendpunct}{\mcitedefaultseppunct}\relax
\EndOfBibitem
\bibitem[Timoshenko and Gere(1961)]{Timoshenko1961}
S.~P. Timoshenko and J.~M. Gere, \emph{{Theory of elastic stability}},
  McGraw-Hill, New York, 2nd edn., 1961\relax
\mciteBstWouldAddEndPuncttrue
\mciteSetBstMidEndSepPunct{\mcitedefaultmidpunct}
{\mcitedefaultendpunct}{\mcitedefaultseppunct}\relax
\EndOfBibitem
\bibitem[Pauchard and Couder(2004)]{Pauchard2004}
L.~Pauchard and Y.~Couder, \emph{EPL}, 2004, \textbf{66}, 667--673\relax
\mciteBstWouldAddEndPuncttrue
\mciteSetBstMidEndSepPunct{\mcitedefaultmidpunct}
{\mcitedefaultendpunct}{\mcitedefaultseppunct}\relax
\EndOfBibitem
\bibitem[Datta \emph{et~al.}(2010)Datta, Shum, and Weitz]{Datta2010}
S.~S. Datta, H.~C. Shum and D.~A. Weitz, \emph{Langmuir}, 2010, \textbf{26},
  18612--6\relax
\mciteBstWouldAddEndPuncttrue
\mciteSetBstMidEndSepPunct{\mcitedefaultmidpunct}
{\mcitedefaultendpunct}{\mcitedefaultseppunct}\relax
\EndOfBibitem
\bibitem[Pozrikidis(2003)]{Pozrikidis2003}
C.~Pozrikidis, \emph{{Modeling and Simulation of Capsules and Biological
  Cells}}, Chapman and Hall/CRC, 2003\relax
\mciteBstWouldAddEndPuncttrue
\mciteSetBstMidEndSepPunct{\mcitedefaultmidpunct}
{\mcitedefaultendpunct}{\mcitedefaultseppunct}\relax
\EndOfBibitem
\bibitem[Libai and Simmonds(1998)]{Libai1998}
A.~Libai and J.~G. Simmonds, \emph{{The Nonlinear Theory of Elastic Shells}},
  Cambridge University Press, 1998\relax
\mciteBstWouldAddEndPuncttrue
\mciteSetBstMidEndSepPunct{\mcitedefaultmidpunct}
{\mcitedefaultendpunct}{\mcitedefaultseppunct}\relax
\EndOfBibitem
\bibitem[Maddocks(1987)]{Maddocks1987}
J.~H. Maddocks, \emph{Arch. Rat. Mech. Anal.}, 1987, \textbf{99}, 301\relax
\mciteBstWouldAddEndPuncttrue
\mciteSetBstMidEndSepPunct{\mcitedefaultmidpunct}
{\mcitedefaultendpunct}{\mcitedefaultseppunct}\relax
\EndOfBibitem
\bibitem[Knoche and Kierfeld(2014)]{Knoche2014}
S.~Knoche and J.~Kierfeld, \emph{EPL}, 2014, \textbf{106}, 24004\relax
\mciteBstWouldAddEndPuncttrue
\mciteSetBstMidEndSepPunct{\mcitedefaultmidpunct}
{\mcitedefaultendpunct}{\mcitedefaultseppunct}\relax
\EndOfBibitem
\bibitem[Knoche and Kierfeld(2014)]{Knoche2014long}
S.~Knoche and J.~Kierfeld, \emph{Eur. Phys. J. E}, 2014, \textbf{37}, 62\relax
\mciteBstWouldAddEndPuncttrue
\mciteSetBstMidEndSepPunct{\mcitedefaultmidpunct}
{\mcitedefaultendpunct}{\mcitedefaultseppunct}\relax
\EndOfBibitem
\bibitem[Quilliet(2012)]{Quilliet2012}
C.~Quilliet, \emph{Eur. Phys. J. E}, 2012, \textbf{35}, 48\relax
\mciteBstWouldAddEndPuncttrue
\mciteSetBstMidEndSepPunct{\mcitedefaultmidpunct}
{\mcitedefaultendpunct}{\mcitedefaultseppunct}\relax
\EndOfBibitem
\bibitem[Vliegenthart and Gompper(2011)]{Vliegenthart2011}
G.~A. Vliegenthart and G.~Gompper, \emph{New J. Phys.}, 2011, \textbf{13},
  045020\relax
\mciteBstWouldAddEndPuncttrue
\mciteSetBstMidEndSepPunct{\mcitedefaultmidpunct}
{\mcitedefaultendpunct}{\mcitedefaultseppunct}\relax
\EndOfBibitem
\bibitem[Callen(1960)]{Callen1960}
H.~B. Callen, \emph{{Thermodynamics}}, John Wiley \& Sons, New York, 1960\relax
\mciteBstWouldAddEndPuncttrue
\mciteSetBstMidEndSepPunct{\mcitedefaultmidpunct}
{\mcitedefaultendpunct}{\mcitedefaultseppunct}\relax
\EndOfBibitem
\bibitem[Lipowsky \emph{et~al.}(2005)Lipowsky, Brinkmann, Dimova, Franke,
  Kierfeld, and Zhang]{Lipowsky2005}
R.~Lipowsky, M.~Brinkmann, R.~Dimova, T.~Franke, J.~Kierfeld and X.~Zhang,
  \emph{J. Phys.: Condens. Matter}, 2005, \textbf{17}, S537--S558\relax
\mciteBstWouldAddEndPuncttrue
\mciteSetBstMidEndSepPunct{\mcitedefaultmidpunct}
{\mcitedefaultendpunct}{\mcitedefaultseppunct}\relax
\EndOfBibitem
\bibitem[Paulose \emph{et~al.}(2012)Paulose, Vliegenthart, Gompper, and
  Nelson]{Paulose2012}
J.~Paulose, G.~A. Vliegenthart, G.~Gompper and D.~R. Nelson, \emph{Proc. Natl.
  Acad. Sci. U. S. A.}, 2012, \textbf{109}, 19551--6\relax
\mciteBstWouldAddEndPuncttrue
\mciteSetBstMidEndSepPunct{\mcitedefaultmidpunct}
{\mcitedefaultendpunct}{\mcitedefaultseppunct}\relax
\EndOfBibitem
\bibitem[Pelekasis and Lytra(2014)]{Pelekasis2014}
N.~Pelekasis and A.~Lytra, \emph{Fluid Dyn. Res.}, 2014, \textbf{46},
  041422\relax
\mciteBstWouldAddEndPuncttrue
\mciteSetBstMidEndSepPunct{\mcitedefaultmidpunct}
{\mcitedefaultendpunct}{\mcitedefaultseppunct}\relax
\EndOfBibitem
\bibitem[Pogorelov(1988)]{Pogorelov1988}
A.~V. Pogorelov, \emph{{Bendings of Surfaces and Stability of Shells}},
  American Mathematical Society, 1988, p.~77\relax
\mciteBstWouldAddEndPuncttrue
\mciteSetBstMidEndSepPunct{\mcitedefaultmidpunct}
{\mcitedefaultendpunct}{\mcitedefaultseppunct}\relax
\EndOfBibitem
\bibitem[Carlson \emph{et~al.}(1967)Carlson, Sendelbeck, and Hoff]{Carlson1967}
R.~L. Carlson, R.~L. Sendelbeck and N.~J. Hoff, \emph{Exp. Mech.}, 1967,
  281--288\relax
\mciteBstWouldAddEndPuncttrue
\mciteSetBstMidEndSepPunct{\mcitedefaultmidpunct}
{\mcitedefaultendpunct}{\mcitedefaultseppunct}\relax
\EndOfBibitem
\end{mcitethebibliography}
\bibliographystyle{rsc} 
}

\end{document}